\documentclass[sigconf]{acmart}

\usepackage{booktabs} 
\usepackage{subfigure}

\usepackage{tabularx}

\setcopyright{rightsretained}

\acmConference[Unreviewed]{}{2018}{Portugal}

\begin{document}
\title{The Bumpy Road of Taking Automated Debugging to Industry}

\author{Rui Abreu}
\orcid{1234-5678-9012}
\affiliation{%
  \institution{IST, University of Lisbon and INESC-ID}
  \country{Portugal}
}
\email{rui@computer.org}

\begin{abstract}
Debugging is arguably among the most difficult and extremely time consuming
tasks of the software development life cycle. Therefore, it comes as no surprise
that researchers have invested a considerable amount of effort in developing
automated techniques and tools to support developers excel in these tasks.
Despite the significant advances, including demonstrations of usefulness, efficacy,
and efficiency, these techniques are yet to find their way into industrial adoption.
In this paper, we reflect upon the commercialization efforts of a particular
automated debugging technique and lay down potential reasons for lack of success
stories as well as ideas to move forward.
\end{abstract}

%
%



\maketitle

\section{Introduction}
Debugging is known to be time consuming and tedious task: debugging is the ugly
duckling of the software development process; yet every developer is destined to
spend significant effort performing this task. The challenge is to find bugs as
early and quickly as possible so that software is shipped without defects.

The research community has embraced this challenge and proposed several approaches
and techniques to support developers while debugging code
bases~\cite{DBLP:journals/tse/WongGLAW16}. Despite these advances to the
state-of-the-art of debugging techniques, there are still no success stories
on the world-wide adoption of these techniques by industry.

In fact, there was a recent study on the the dichotomy of debugging behavior among
programmers~\cite{beller2018dichotomy} were it was confirmed that IDE-based
debuggers are not used as often as the so-called ``printf debugging''. The authors
further of this study mention that their results ``call to strengthen hands-on
debugging experience in computer science curricula and have already refined the
implementation of modern IDE debuggers''.

We have recently embarked in an effort to commercialize a particular automated
debugging technique. In the next sections, we outline the automated debugging
technique we attempted to commercialize and reflect upon the challenges that
prevented successful commercialization. Finally, we conclude with suggestions
for the software engineering community to move forward.
\section{Automated Debugging}
This section discusses the most common automated debugging approaches and
a materialization into a toolset of a promising and lightweight technique.
\subsection{Debugging Approaches}
Debugging techniques exploit information obtained from:

\begin{itemize}
  \item[(i)] the execution of the program;
  \item[(ii)] analysis of the behavior of a program or its model; and
  \item[(iii)] from given specifications.
\end{itemize}

\noindent Automation is essential as the size of programs and obtained
information precludes manual investigation, yet different debugging methods
remain isolated from each other~\cite{DBLP:journals/tse/WongGLAW16}.

Statistics-based techniques (e.g.,~\cite{DBLP:conf/icse/PearsonCJFAEPK17})
are popular automated fault localization techniques within the Software Engineering
community. They correlate information about program fragments that have been
exercised in multiple program execution traces (also called \textit{program spectra})
with information about successful and failing executions. By doing that,
spectrum-based fault localization (SFL) and other statistics-based approaches
yield a list of suspect program fragments sorted by their likelihood to be at
fault. Since this technique is efficient in practice, it is attractive for large
modern software systems~\cite{ecbs07}.

Despite the efforts to advance SFL to efficiently aid programmers to pinpoint the
root cause of observed failures, it still has a few drawbacks:

\begin{itemize}
  \item Spectrum-based fault localization only exploits the topology of the
  software. Thus, low granularity components (e.g., statement level) will yield
  the best diagnostic performance, whereas coarser grain granularity may guide
  the developer to inspect healthy components.

  \item Due to its statistical foundation, the diagnostic accuracy is inherently
  limited: (1) the accuracy is rather dependent on the number and quality of the
  test cases, (2) it cannot reason over multiple faults, like model-based
  diagnosis approaches.

  \item The similarities are not probabilities: this hampers the analysis of the
  ranking using AI, probabilistic methods.
\end{itemize}

Apart from the aforementioned issues, such lightweight techniques are not yet
able to consistently pinpoint a fault, if used in isolation~\cite{ase09}.
Although SFL methods can identify promising starting points for investigation,
further filtering, symptom-based aggregation, model-based reasoning and guidance
mechanisms are desired~\cite{Parnin:2011:ADT:2001420.2001445,DBLP:conf/kbse/MayerS08}.
Furthermore, multiple simultaneous faults affecting the program are common yet
poorly addressed~\cite{debroy2011towards,ase09}.

%

Despite the aforementioned limitations, we argue that current state-of-the-art
approaches carries valuable information to warrant being transferred to industry.
In particular, there are materializations of spectrum-based fault localization
that are \textit{production} ready. An example is \textsc{GZoltar} -- a technique
we will outline next.
\subsection{The GZoltar toolset}
\textsc{GZoltar}~\cite{DBLP:conf/kbse/CamposRPA12} is an Eclipse plugin~\cite{Burnette:2005:EIP:1199184}
for automatic testing and debugging. Currently, the toolset is provided as a
plugin and library, and integrates seamlessly with the JUnit framework.
It is an evolution from Zoltar~\cite{Zoltar} (addressing the C programming
language) and is implemented using Java and analyses programs written in Java.
To install the \textsc{GZoltar} toolset, users need to request a license at
\texttt{http://www.gzoltar.com/}.

\begin{figure}
  \includegraphics[width=0.25\textwidth,height=0.23\textheight]{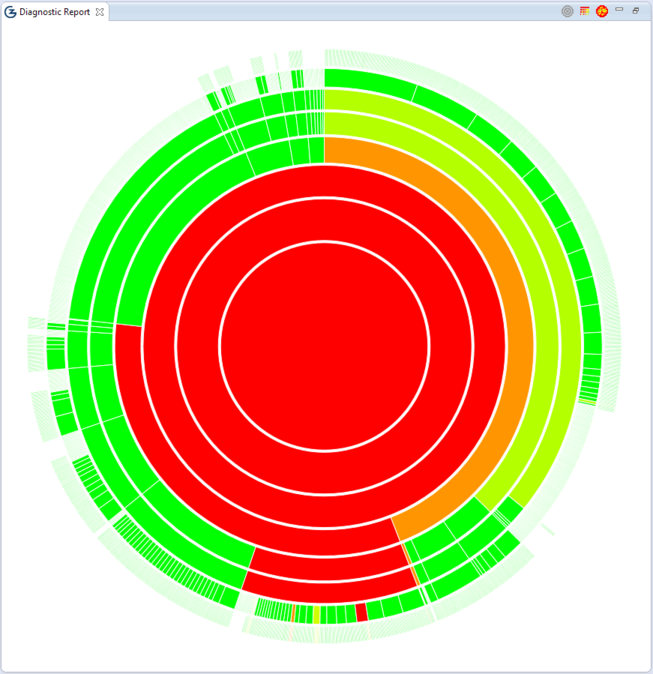}
  \caption{GZoltar's Sunburst Visualization}
  \label{fig:sunburst}
\end{figure}

As said before, currently \textsc{GZoltar} offers three distinct visualizations
implemented in HTML5: Sunburst (see Figure~\ref{fig:sunburst} as an example), Vertical
Partition and Bubble Hierarchy. We only show the sunburst visualization due to space
limitations; interested readers can refer the the official website to see the other
visualizations. Sunburst is also the one preferred by the users~\cite{DBLP:conf/vissoft/GouveiaCA13}.
Note that this toolset uses a visualization framework, D3.js, which allows
the creation of new visualizations with little effort. D3.js is a JS library which
allows the creation of different visual representations of data.

The generated visualizations are interactive, and the user is able to navigate
through the project structure to analyze it in detail.
The main goal of the visualizations are to represent the analyzed project in an
hierarchical way to allow a faster and easier debugging process. For instance,
in the Sunburst visualization, each ring denotes an hierarchical level of the
source code organization (from the inner to the outer circle).

All visualizations obey to a color gradient ranging from green (low suspiciousness)
to red (very high suspiciousness). The suspiciousness is computed by the diagnostic
algorithm detailed in~\cite{DBLP:conf/kbse/CamposRPA12}.
If all the tests passed (i.e., there were no observed failures), the underlying
fault localization technique yields an empty diagnostic report. As such,
\textsc{GZoltar} will render a visualization with all the system components in green.
It should be noted that this does not mean that the system under test is bug-free,
but rather that no failure was observed.

Users are able to navigate and interact with the visualizations.
They may analyze each component by hovering the mouse cursor at any element on
any visualization and a tooltip is shown with the identification of it and
the corresponding suspiciousness value.
The user may click on any component representation to automatically open the
code editor with the respective source code.
Users may also zoom in/out the visualization to analyze
in detail a specific part of the system, and pan the visualization.
The user may also resize the diagnostic report view at any time to automatically
enlarge or reduce the visualization used. There are other navigation features
available, but due to space limitations we refrain from explaining them.

The commercial efforts of \textsc{GZoltar}, discussed next, were coined Crowbar
instead.
\section{Crowbar}
As we developed the underline algorithms and techniques in the context of research
projects funded by the Portuguese Foundation for Science and Technology (FCT) and
the National Science Foundation (NSF), we felt the urge to transfer the technology
into industry. In fact, the idea of transferring a toolset offering state-of-the-art
automated debugging capabilities was something that came to mind during the course
of the TRADER project\footnote{https://bit.ly/2JOUBe4}. The objective of the TRADER
project, funded by the Dutch Ministry of Economic Affairs, was to develop methods and
tools for ensuring reliability of consumer electronic products. The objective was
to minimize product failures that are exposed to the end user. The proposed techniques
were evaluated using use cases from Philips Semiconductors' Innovation Center
Eindhoven in the area of digital television.

We assembled a team consisting of a Head of Engineering, developers (mostly PhD
students), and a business developer (at the time a final-year master student in the
Master in Innovation and Technological Entrepreneurship at the University of
Porto\footnote{https://bit.ly/2LNqxwO}). To get the ball rolling, we started by
applying to UPTEC's\footnote{UPTEC - Science and Technology Park of University of
Porto supports the creation of technological, scientific and creative based
companies and attracts national and international innovation centers.}
Startup Acceleration Program. The idea of this program is to to raise awareness
among new entrepreneurs about the complex challenges in the process of creating
and developing a business project. The Acceleration Program culminated with a
Startup Pitch Day -- which we won~\footnote{https://bit.ly/2Ml2lmu}.

Winning the Startup Day helped to raise awareness and gave us momentum. As the
project started to gain traction, we were invited to TV programs and were featured
in specialized magazines.  More importantly, though, companies started to reach
out to us to better understand our technology and offer themselves to help in the
industrial validation of the tool.

The industrial validation, however, proved to be extremely challenging. Finding
industrial partners to try our tools out was not as difficult as one would imagine,
but having the tools bullet proof to all sorts of situations was far from a
reality.
Due to confidentiality
reasons, we cannot name our industrial partners. Rest assured that the industrial
partners came from all industrial domains: from aerospace to banking and automotive
industries.

Regardless of the technological challenges, we have engaged in discussions
with a Venture Capital firm (VC). Once again, due to confidentiality reasons,
the VC cannot be named. The VC has valued our prototype in 2 million euro,
and they would be willing to invest 500k euro. However, we did not reach an
agreement because the milestones were not realistic and the VC was not willing
to take the necessary risk. This was a considerable blow in our efforts.
\section{Challenges}
In our efforts to incorporate a startup to offer automated debugging techniques
we have faced several hurdles. We next discuss these challenges and outline
potential solutions.

\begin{itemize}
  \item The core team was not 100\% committed to the commercialization effort.

  The engineering team was composed mostly of PhD students. This meant that there
  were some opposing forces which were difficult to manage. On the one hand, the
  students wanted to do a prototype to get results for the next paper, but on the
  other hand our industrial partners wanted mature tools to play with.

  \item The way researchers evaluate their techniques do
  not mimic software developer debugging observed failures.

  Most research on automated fault localization use metrics that completely disregard
  the human factor~\cite{Parnin:2011:ADT:2001420.2001445,DBLP:conf/vissoft/GouveiaCA13}.
  For instance, such metrics
  assume perfect bug understanding by developers and
  that developers would religiously follow the ranked list of suspicious statements
  given by the automated fault
  localization tool. This is just not right~\cite{ecbs07}. In fact, extensive user
  studies are needed to understand whether diagnostic reports generated by current
  state-of-the-art techniques are indeed able to convey actionable insights
  once the human factor is considered. User studies require considerable effort
  and it might the a good idea to create a world-wide consortium of researchers and
  people from industry to carry a comprehensive user study to understand where we
  stand.

  \item Top-level managers do not see the added value of paying a debugging technique.

  In most conversations with several industrial partners (both in Portugal and in
  the Silicon Valley) a common comment came up: as a top-level manager I only see
  code that is ready to deploy. In other words, despite the several stories of
  software failures, top-level managers assume that this is a problem that is
  not going to affect them because they have hired the most stellar software engineers.
  However, as you can imagine, even the most stellar software engineers commit mistakes.
  We need a novel way to quantify the gains for the business of having automated
  debugging techniques helping the debugging process.

  \item Code instrumentation challenges.

  Our team faced several challenges maturing our code instrumentation toolset to
  gather the test-coverage information required for our diagnostic analysis. Although
  we managed to improve the toolset at a considerably fast pace, our industrial partners
  would frequently face issues regarding the instrumentation. In several cases, even after
  signing non-disclosure agreements, we were unable to gather and collect metrics that
  would help us pinpoint the root cause, much less were we able inspect the source code
  that caused such instrumentation to fail. Interestingly, we have ensured that our tools
  -- which were initially developed in an academic context -- performed reliably in hundreds
  of open-sourced projects (as evidenced by the breadth of projects the empirical evaluations
  cover~\cite{DBLP:conf/icse/PearsonCJFAEPK17,DBLP:conf/icst/PerezAd17}). This suggests
  that there is some disconnect between open-sourced software and its close-sourced counterpart,
  at least regarding the use language features and the choice of architectures, impacting
  the way code coverage needs to be instrumented, orchestrated, and gathered.

  \item Test Cases: The Holy Grail.

  Most companies do not have a systematic way to test their apps. This becomes
  a challenge as SFL is a dynamic technique that requires the system to be
  exercised with a diverse set of cases. If test cases do not exist, then SFL-like
  techniques will not be of any help. This challenge highlights the need to offer
  integrated solutions to automated test case generation and debugging. One of the
  major challenges is to solve the oracle problem, as there are independent techniques
  for both tasks. For instance, EvoSuite~\cite{DBLP:conf/sigsoft/FraserA11} for
  test case generation and \textsc{GZoltar} for debugging.
  A technique that is able to automatically assert whether a test case passes or
  fails will have a tremendous contributions in making techniques such as \textsc{GZoltar}
  being mass-adopted in industry.

  Yet another challenge related to test cases is that only recently the community
  started to investigate what properties of a test suite make it suitable for
  automated debugging~\cite{DBLP:conf/icse/PerezAD17}. This understanding is however
  crucial if one plans to use automated debugging techniques effectively. Recently,
  we proposed a metric, DDU~\cite{DBLP:conf/icse/PerezAD17}, that quantifies the
  diagnosability of a test suite. This metric evaluates several interesting properties of a test suite regarding its ability to isolate faults
  and has the potential to either replace or co-exist with the more common coverage metrics.
  DDU information helps to rethink how to evolve the test suite: e.g., adding more
  unit tests or system tests. Yet another added value is that DDU can be used to guide
  automated test generation techniques, such as the work
  in~\cite{DBLP:conf/kbse/CamposAFd13,campos2014continuous}.
  However to foster adoption of DDU -- and hence automated
  debugging techniques -- researchers still need to find a way to convey that metric
  to the developer (again, explainability is a concern). Also, there are no studies as to whether the
  metric has the capability of quantifying the testability of a test suite.

  \item Diagnostic Reports: Interpretation and Expectations.

  Most fault localization techniques output a ranked list of components, possibly with a
  probability of that component be the true faulty component. As the developer
  looks into the ranking, there is no information as to why a given statement
  is considered to be potentially faulty. Furthermore, there is also no contextual information
  about observations in failed and passed executions in the diagnostic report. In other words, one is offered
  no explanation that would help understand the ranked list of suspicious components. An approach was recently
  proposed~\cite{DBLP:conf/ijcai/PerezA18} to augment the ranking with qualitative, contextual
  information as an attempt to explain the ranking. This work lays the first stone in a series
  of efforts to more deeply integrate reasoning-based Artificial Intelligence approaches into
  SFL. Notably, it paves the way for further efforts by the fault localization
  research community, namely by: (1) Improving automated landmarking by expanding
  its application to complex non-primitive objects and by exploring ensembles of
  multiple strategies; (2) Conducting a systematic user study investigating the
  extent that qualitative domain partitioning aids bug understanding.

  Another challenge that needs to be addressed is the fact that developers need
  to be educated on the capabilities of the tool as to set proper expectations.
  The tool is just an aid to help developers debug failures and one should take
  the diagnostic reports: there might be false positives ranking as high as
  faulty statements. Addressing this challenge would require to include such
  techniques in course work of undergraduate curricula.

  \item Multi-platform and development environments.

  The current version of the tool is offered as an Eclipse IDE only. This hinders
  adoption by industrial partners. Support to other environments would potentially
  increase wide adoption. An idea that comes to mind is to integrate the toolset
  into the CI/CD pipeline -- an analysis will be performed when a pull request is
  made and the developer would get the visual diagnostic report as a comment to
  the pull request is something fails.
\end{itemize}
\section{Conclusions}
Transferring automated fault localization techniques to industry has been far from
trivial. Given the challenges outlined in the previous section, we urge the
software engineering community to create an interest group, including groups in
academia and industry, to publicly promote and create advantages seeking wide
adoption of automated debugging techniques. The main goal of such interest group
would be to develop an open source platform offering automated testing and fault
localization techniques.

As starting point, we would suggest to extend current standard debugging
techniques (e.g. breakpoints) of IDEs, such as Jetbrain's IntelliJ and Microsoft
Visual Studio Code, with automated debugging tools. This is aligned with the most
common feedback we obtained: gain traction by offering these techniques publicly
on commonly used IDEs. Without the
tool vendors on board, it becomes extremely difficult to succeed.

To encourage developers and companies alike to voluntarily participate in this
interest group, we further suggest that the \textit{modus operandi} of this group
to follow the free and open-source software principles: so that anyone is freely
licensed to use, copy, study, and change the software in any way, and the source
code is openly shared so that people are encouraged to contribute. The interest
group should further seek a sponsorship model among all participants.

We are hopeful that this paper lays the first stone towards this interest group
and that next steps are discussed at the conference.

\begin{quote}
\textit{Audere est facere}.
\end{quote}

\bibliographystyle{ACM-Reference-Format}
\bibliography{paper}

\end{document}